\documentclass[twocolumn,showpacs,preprintnumbers]{revtex4}

\usepackage{amsfonts,amssymb,amsmath,amsthm}

\newtheorem{theorem}{Theorem}

\newtheorem{cor}{Corollary}
\newtheorem{lemma}{Lemma}
\newtheorem{dfn}{Definition}

\newcommand{\RR}{\mathbb{R}}

\newcommand{\CC}{\mathbb{C}}

\newcommand{\LL}{\mathbb{L}}

\newcommand{\calH}{{\cal H }}

\newcommand{\la}{\langle}
\newcommand{\ra}{\rangle}

\newcommand{\rk}[1]{\mathop{\mathrm{Rk}(#1)}}
\newcommand{\range}[1]{\mathop{\mathrm{Rg}(#1)}}
\newcommand{\kernel}[1]{\mathop{\mathrm{Ker}(#1)}}
\newcommand{\image}[1]{\mathop{\mathrm{Im}(#1)}}
\newcommand{\ep}{\epsilon}

\newcommand{\be}{\begin{equation}}
\newcommand{\ee}{\end{equation}}
\newcommand{\beea}{\begin{eqnarray}}
\newcommand{\eea}{\end{eqnarray}}
\newcommand{\nn}{\nonumber}
\newcommand{\ba}{\begin{array}}
\newcommand{\ea}{\end{array}}

\usepackage{graphicx}
\usepackage{dcolumn}
\usepackage{bm}

\DeclareMathOperator*{\tr}{\mathop{\mathrm{Tr}}}

\begin{document}

\title{Unextendible product bases and locally unconvertible
bound entangled states}

\author{Sergey Bravyi}
\email{serg@cs.caltech.edu}
\affiliation{Institute for Quantum Information,
California Institute of Technology,\\
Pasadena, 91125 CA, USA.}

\date{\today}

\begin{abstract}
Mutual convertibility of bound entangled states under local
quantum operations and classical communication (LOCC) 
is studied. We focus on states associated with unextendible
product bases (UPB) in a system of three qubits. A complete 
classification of such UPBs is suggested. We prove that for
any pair of UPBs $S$ and $T$ the associated bound entangled 
states $\rho_S$ and $\rho_T$ can not
be converted to each other by LOCC, unless $S$ and $T$ 
coincide up to local unitaries. 
More specifically, there exists a
finite precision $\ep(S,T)>0$ such that for any LOCC protocol
mapping $\rho_S$ into a probabilistic ensemble $(p_\alpha,\rho_\alpha)$,
the fidelity between $\rho_T$ and any possible
final state $\rho_\alpha$
satisfies $F(\rho_T,\rho_\alpha)\le 1-\ep(S,T)$.
\end{abstract}

\pacs{03.65.Bz, 03.67.-a, 89.70.+c}

\maketitle

\section{\label{sec:intro} Introduction and summary of results}
One of the most challenging problems in the field of quantum information 
is to understand what transformations of multipartite entangled states
can be realized by local quantum operations and classical communication (LOCC).
This problem has many important applications such as
construction of entanglement distillation
protocols~\cite{BBPSSW96}, quantum data hiding~\cite{DLT01}, and
separability check~\cite{Werner89}. 
The following question is of particular importance for
all these applications. Given mixed states $\rho_S$ and $\rho_T$ of
some fixed composite quantum system, is it possible to
convert $\rho_S$ into $\rho_T$ by LOCC transformations?
More specifically, we will be interested in stochastic 
approximate convertibility. 
Recall, that an output of a LOCC transformation 
applied to the state $\rho_S$
is a probabilistic ensemble $(p_\alpha,\rho_\alpha)$, where $\alpha$ is
a classical random variable describing results of intermediate 
measurements and having
a probability distribution $\{p_\alpha\}$.
The state $\rho_\alpha$ is an output quantum state conditioned on $\alpha$.
By {\it stochastic convertibility} we
mean that $\rho_\alpha$ must coincide with $\rho_T$ for 
at least one outcome $\alpha$ (we assume that $p_\alpha>0$
for all $\alpha$).  In other words, the conversion succeeds with
some probability which is generally smaller than one. 
By {\it stochastic approximate convertibility} we mean that for
any given precision $\ep>0$ the state $\rho_S$ can be
stochastically converted into some state $\rho_\ep$ satisfying
$F(\rho_T,\rho_\ep)\ge 1-\ep$, where $F$ is the fidelity.
Note that the probability for the conversion to succeed may turn to zero
as the precision $\ep$ turns to zero. If this is the case, 
we can not stochastically convert $\rho_S$ into $\rho_T$.
The stochastic approximate convertibility seems to be the
weakest one as far as one-copy-to-one-copy transformations are
concerned.

For last several years 
the stochastic convertibility of  multipartite {\it pure} states 
has been extensively studied.
In the case of pure states LOCC transformation reduces to local filtering
operations~\footnote{By local
filtering operations we mean a transformation $\rho\to X\rho X^\dag$,
where $X=X_A\otimes X_B\otimes\cdots$ is an arbitrary product
operator.}.
 The question of convertibility by local filtering operations (LFO)
can be investigated using the theory of normal forms
and entanglement monotones which are invariant under LFO,
see Ref.~\cite{VDD01}. 

Conditions for stochastic convertibility of mixed states are
less understood. It was conjectured in Ref.~\cite{VDD02}
that a two-qubit mixed state $\rho_S$ can be stochastically converted
into a state $\rho_T$ iff the Bell-diagonal  normal form of $\rho_T$
equals to convex sum of some separable state and
the Bell-diagonal  normal form of $\rho_S$. 

In this paper we consider only mixed states of a very special form,
namely bound entangled states associated with unextendible 
product bases in a system of three qubits. The notion of unextendible
product basis (UPB) was originally introduced in Ref.~\cite{BDMSST99}.
In our setting UPB is a family of vectors $|S_1\ra,\ldots,|S_n\ra
\in (\CC^2)^{\otimes 3}$ such that
\begin{itemize}
\item Each vector $|S_j\ra$ has a product form
$|S_j\ra=|A_j\ra\otimes |B_j\ra\otimes |C_j\ra$ for some
one-qubit states $|A_j\ra$, $|B_j\ra$, and $|C_j\ra$.
\item  $\la S_i|S_j\ra=\delta_{ij}$ for all $i$ and $j$,
\item The orthogonal complement to the space spanned by 
the vectors $|S_1\ra,\ldots,|S_n\ra$ does not contain
product vectors.
\end{itemize}
The vectors $|S_j\ra$ are refered to as members of the UPB.
If $S=\{|S_j\ra\}$ is a UPB, the linear space spaned by its members will
be denoted $\calH_S$. A mixed state
$\rho_S$ associated with a UPB $S$ is defined as a properly normalized 
projector onto an orthogonal complement of $\calH_S$.
In our setting it is
\be\label{rhoS}
\rho_S = \frac1{2^3-n} \left(I-\sum_{j=1}^n |S_j\ra\la S_j|\right).
\ee
It can not be a separable state, since its range does not contain
product vectors. However, as was pointed out in Ref.~\cite{DMSST00},
in the case of three qubits $\rho_S$ is separable with respect
to any {\it bipartite} cut,
for instance $ABC=A\cup BC$ (we will use letters A, B, and C to
label the qubits).
In particular it is not possible
to distill some pure entanglement between any two qubits
starting from many copies of $\rho_S$ and applying only
LOCC transformations. For that reason $\rho_S$ is refered
to as a bound entangled state.

In section~\ref{sec:upb} we build a complete classification of
three-qubit UPBs. It appears that each UPB has exactly four members.
By local unitaries and permutations of the members
we can bring any UPB into the following form:
\beea\label{intro1}
|S_1\ra&=&|0\ra\otimes|0\ra\otimes|0\ra,\nn \\
|S_2\ra&=&|1\ra\otimes|B\ra\otimes|C\ra,\nn \\
|S_3\ra&=&|A\ra\otimes|1\ra\otimes|C^\perp\ra,\nn \\
|S_4\ra&=&|A^\perp\ra\otimes|B^\perp\ra\otimes|1\ra,
\eea
where
\beea\label{intro2}
|A\ra\la A| &= & (1/2)\left[I+ \cos{(\theta_A)}\sigma^z +
\sin{(\theta_A)}\sigma^x\right],
\nn \\
|B\ra\la B| &= & (1/2)\left[I+ \cos{(\theta_B)}\sigma^z +
\sin{(\theta_B)}\sigma^x\right],
\nn \\
|C\ra\la C| &= & (1/2)\left[I+ \cos{(\theta_C)}\sigma^z +
\sin{(\theta_C)}\sigma^x\right].
\eea
(Here and throught the paper we use a designation $|A^\perp\ra$
for a state orthogonal to the  state $|A\ra$.) 
The family given by Eq.~(\ref{intro1},\ref{intro2}) specifies a UPB
for an arbitrary triple $\theta_A,\theta_B,\theta_C \ne 0 \mod{\pi}$.
However some triples must be identified since the corresponding
UPBs can be matched by local unitaries and permutation of the 
UPB's members (in this case the associated states are related by local
unitaries).
We show that the "fundamental" region 
of parameters corresponds to $\theta_A,\theta_B,\theta_C \in (0,\pi)$.
Restricting ourselves to this region we count each UPB exactly
one time.

Our main conclusion is that stochastic approximate LOCC conversion
(in either direction)  of states
$\rho_S$ and $\rho_T$ associated
with UPBs $S$ and $T$ 
is impossible, unless $\rho_S$ and $\rho_T$
are related by local unitaries. Following Ref.~\cite{Rains97}
we obtain the necessary conditions for LOCC convertibility 
finding the necessary conditions for convertibility
by separable superoperators. So the statement which we
have actually proved is following:
\begin{theorem}\label{theorem:main}
Suppose $S$ and $T$ are three-qubit UPBs which are not
related by local unitary operators and permutation of the
members. Let $\rho_S$ and $\rho_T$ be the bound entangled
states associated with $S$ and $T$. There exists a finite
precision $\ep(S,T)>0$ such that for any 
separable completely positive superoperator $E$ 
satisfying $E(\rho_S)\ne 0$ one has
\[
F\left(\rho_T,\, \frac{E(\rho_S)}{\tr[E(\rho_S)]}\right) \le 1-\ep(S,T).
\]
\end{theorem}

The paper is organized in the following way. In section~\ref{sec:upb}
we build a classification of three-qubit UPBs and reveal some
useful facts about them. The most important fact is 
Lemma~\ref{lemma:no-product-vectors} which says that members
of UPB are the only product vectors in the spanning space of a UPB.
In section~\ref{sec:main} we address stochastic (exact) convertibility
of mixed states associated with UPBs.
In section~\ref{sec:approx1} we examine a simplified version of
the problem, namely approximate convertibility by local filtering
operations. Section~\ref{sec:approx2} contains a proof
of Theorem~\ref{theorem:main}. In Conclusion we summarize
the results obtained in the paper and discuss possible application
of our method to UPBs in a system $\CC^3\otimes \CC^3$.

\section{\label{sec:upb} Unextendible product bases for three qubits}
In this section we put forward a complete classification of 
UPBs for a system of three qubits and prove some useful facts
about them. The qubits will be refered to as A, B, C.
Let $|S_j\ra\in \CC^2\otimes \CC^2\otimes \CC^2$ be members of UPB:
\be\label{common-UPB}
|S_j\ra = |A_j\ra\otimes |B_j\ra \otimes |C_j\ra, \quad
j=1,\ldots,n.
\ee
By definition $\la S_i|S_j\ra=\delta_{i,j}$ for all $i$ and $j$.
Let $\calH_S$ be a linear
$n$-dimensional space spanned by the basis vectors $S_j$ and 
$\rho_S$ be a density operator proportional to a projector onto
orthogonal complement of $\calH_S$, see Eq.~(\ref{rhoS}).
As an example consider a UPB "Shifts" suggested in Ref.~\cite{BDMSST99}.
Its members are 
\beea
|S_1\ra&=&|0\ra\otimes|0\ra\otimes|0\ra, \nn \\
|S_2\ra&=&|1\ra\otimes|-\ra\otimes|+\ra, \nn \\
|S_3\ra&=&|+\ra\otimes|1\ra\otimes|-\ra, \nn \\
|S_4\ra&=&|-\ra\otimes|+\ra\otimes|1\ra. \nn 
\eea
(Here we use a standard notations $|\pm\ra=2^{-1/2}(|0\ra\pm |1\ra)$.)
If some product vector $|S\ra=|A\ra\otimes|B\ra\otimes|C\ra$
extends this basis, then one of vectors
$|A\ra$, $|B\ra$, $|C\ra$ must be orthogonal to at least
two vectors from the family $\{ |0\ra, |1\ra, |+\ra, |-\ra\}$,
which is impossible.

An important characteristic of an UPB is its 
orthogonality graph, see Ref.~\cite{DMSST00}. For our purposes it will be
convenient to introduce more general definition of an
orthogonality graph.
\begin{dfn}
Let $M=\{|\Psi_1\ra,\ldots,|\Psi_N\ra\in \calH\}$ be a family
of vectors in a finite-dimensional Hilbert space $\calH$.
An orthogonality graph $G=(V,E)$ of the family $M$
has a set of vertices $V=\{1,\ldots,N\}$ with one vertex 
assigned to each  state $|\Psi_j\ra$. A pair of vertices $(i,j)$
is an edge, $(ij)\in E$, iff $\la \Psi_i|\Psi_j\ra=0$. 
\end{dfn}
For example, an orthogonality graph of any family of vectors in
$\CC^2$ is a collection of disjoint edges and isolated vertices.
This definition will help us to prove the following statement.

\begin{lemma}
Any three-qubit UPB has four members.

\begin{proof}
First of all we note that the number of members $n$ can not 
be greater than 5. Indeed, if $n\ge 6$, we have
$\rk{\rho_S}\le 2$. From Ref.~\cite{KL01}
we know that a PPT three-qubit state with a rank two or one
must be separable. But $\rho_S$
is not separable by definition.
Let us prove that an existence of a UPB with five members also
leads to a contradiction. Let $S$ be a such UPB with the members
given by Eq.~(\ref{common-UPB}).
Consider orthogonality graphs $G_A=(V,E_A)$, 
$G_B=(V,E_B)$, and $G_C=(V,E_C)$
for the families $\{|A_j\ra\}$, $\{|B_j\ra\}$, and $\{|C_j\ra\}$
respectively.  Here $V=\{1,\ldots,5\}$ and
\beea
(i,j)\in E_A&\mbox{iff}& \la A_i|A_j\ra =0,\nn\\
(i,j)\in E_B&\mbox{iff}& \la B_i|B_j\ra =0,\nn\\
(i,j)\in E_C&\mbox{iff}& \la C_i|C_j\ra =0,\nn
\eea
Orthogonality of the basis vectors  $\la S_i|S_j\ra=\delta_{i,j}$
implies that
any pair of vertices $(i,j)$ belongs to at least one of the
sets $E_A, E_B, E_C$. Thus 
\[
|E_A|+|E_B|+|E_C|\ge  \frac{n(n-1)}2 = 10
\]
and at least one of the sets $E_A$, $E_B$, $E_C$ contains
4 or more edges. Assume that $|E_A|\ge 4$
and focus on the graph $G_A$. Observe that $G_A$
can not have a vertex with a valence 3 or greater. Indeed,
if, say, $(1,2)\in E_A$,   $(1,3)\in E_A$, and $(1,4)\in E_A$
then a state $|S_6\ra=|A_1\ra\otimes|B_1^\perp\ra\otimes |C_5^\perp\ra$
extends a basis which is impossible.
Besides, $G_A$ can not contain cycles with odd number of edges
(this constraint comes from two-dimensional geometry).
Summarizing, $G_A$ must match the following restrictions:
\begin{itemize}
\item  there are at least 4 edges

\item a valence of any vertex may be 0,1,2 only

\item there are no odd cycles 

\end{itemize}
Up to permutations of the vertices there are only two graphs
satisfying all the restrictions. They are shown on FIG.~1.
\begin{figure}
\includegraphics[scale=0.4]{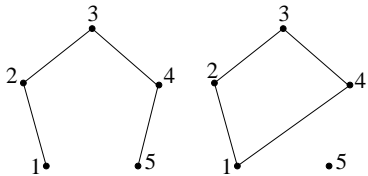}
\caption{Possible orthogonality graphs $G_A$
for the family $\{|A_j\ra\}$.}
\end{figure}
For the graph on the left the basis can be extended by 
a state $|S_6\ra=|A_2\ra\otimes|B_2^\perp\ra\otimes|C_4^\perp\ra$.
For the graph on the right the basis can be extended by
a state $|S_6\ra=|A_5^\perp\ra\otimes|B_5\ra\otimes|C_5\ra$.
Thus an assumption that UPB
contains 5 members leads to a contradiction.
\end{proof}
\end{lemma}

For our purposes the order of UPB members will not be important.
Besides we would like to identify UPBs which can be matched by
local unitary operators. Let us introduce the following equivalence
relation: 
\begin{dfn}\label{def:equiv}
UPBs $S$ and $S'$ are equivalent, $S\sim S'$, iff 
$|S_j\ra=U_A\otimes U_B\otimes U_C|S'_{\sigma(j)}\ra$,
$j=1,2,3,4$,
for some unitary operators $U_A$, $U_B$, $U_C$ and
some permutation $\sigma\in S_4$.
\end{dfn}
We will see later (corollary to 
Lemma~\ref{lemma:no-product-vectors}) that
UPBs $S$ and $S'$ are eqivalent iff the corresponding
states $\rho_S$ and $\rho_{S'}$ are related by local unitaries.
Classification of UPB with 4 members is given by 
\begin{lemma}\label{lemma:UPBs}
Each equivalence class of UPBs has a representative
of the form
\beea\label{canon_UPB}
|S_1\ra&=&|0\ra\otimes|0\ra\otimes|0\ra,\nn \\
|S_2\ra&=&|1\ra\otimes|B\ra\otimes|C\ra,\nn \\
|S_3\ra&=&|A\ra\otimes|1\ra\otimes|C^\perp\ra,\nn \\
|S_4\ra&=&|A^\perp\ra\otimes|B^\perp\ra\otimes|1\ra. 
\eea
For arbitrary  $|A\ra,|B\ra,|C\ra \notin \{ |0\ra,|1\ra\}$
the set of states given by Eq.~(\ref{canon_UPB})
is a UPB.
\end{lemma}
\begin{proof}
Suppose $S=\{|S_j\ra\}$ is a UPB with four members
given by Eq.~(\ref{common-UPB}).
The unextendability implies that  neither of the sets
$\{|A_j\ra\}$, $\{|B_j\ra\}$, $\{|C_j\ra\}$ can contain
two equal states. Indeed, if, say, $|A_1\ra=|A_2\ra$,
then a state $|S_5\ra=|A_1^\perp\ra\otimes|B_3^\perp\ra\otimes|C_4^\perp\ra$
extends the basis. One remains to take into account mutual orthogonality
of the basis members. 
Applying local unitary operators we can make
$|S_1\ra=|0\ra\otimes|0\ra\otimes|0\ra$.
It follows that each of $|S_2\ra$, $|S_3\ra$,
$|S_4\ra$ contains at least one factor $|1\ra$.
Since neither of the sets $\{|A_j\ra\}$, $\{|B_j\ra\}$, $\{|C_j\ra\}$
can contain two copy of $|1\ra$, 
by appropriate permutation of the members
$|S_2\ra$, $|S_3\ra$, $|S_4\ra$ we can always make 
\beea\label{ABC}
|S_1\ra&=&|0\ra\otimes|0\ra\otimes|0\ra,\nn\\
|S_2\ra&=&|1\ra\otimes|B_2\ra\otimes|C_2\ra,\nn\\
|S_3\ra&=&|A_3\ra\otimes|1\ra\otimes|C_3\ra,\nn\\
|S_4\ra&=&|A_4\ra\otimes|B_4\ra\otimes|1\ra,
\eea
Orthogonality $\la S_2|S_3\ra=0$ translates into
$\la C_2|C_3\ra=0$. Analogously, $\la S_2|S_4\ra=0$
implies that  $\la B_2|B_4\ra=0$ and
$\la S_3|S_4\ra=0$ gives us $\la A_3|A_4\ra=0$.
Introducing $|A\ra=|A_3\ra$, $|B\ra=|B_2\ra$, and
$|C\ra=|C_2\ra$ we arrive to a representation
declared in Eq.~(\ref{canon_UPB}).

Consider an orthogonal basis $\{|S_j\ra\}$ as in Eq.~(\ref{canon_UPB})
and suppose that $|A\ra,|B\ra,|C\ra \notin \{ |0\ra,|1\ra\}$.
Assume that a state $|a\ra\otimes|b\ra\otimes|c\ra$ extends
the basis. It is orthogonal to each of $|S_j\ra$, $j=1,2,3,4$
either on the qubit A, or on the qubit B, or on the qubit C.
But each of the states $|a\ra$, $|b\ra$, $|c\ra$
may provide orthogonality to at most one member of
the family $\{|S_j\ra\}$. Thus the basis is unextendible.
\end{proof}

Let us count, how many real parameters we need to parameterize
equivalence classes of  UPBs. Applying local phase shifts
to the qubits A, B, C, we can always make
\beea\label{param1}
|A\ra\la A| &= & (1/2)(I+ \cos{(\theta_A)}\sigma^z + \sin{(\theta_A)}\sigma^x),
\nn \\
|B\ra\la B| &= & (1/2)(I+ \cos{(\theta_B)}\sigma^z + \sin{(\theta_B)}\sigma^x),
\nn \\
|C\ra\la C| &= & (1/2)(I+ \cos{(\theta_C)}\sigma^z + \sin{(\theta_C)}\sigma^x).
\eea
Different triples $(\theta_A,\theta_B,\theta_C)$ may still represent
the same equivalence class. For example, applying $\sigma^z$ to the
qubit A, we should identify $\theta_A$ and $-\theta_A$ (analogously,
$\theta_B\equiv -\theta_B$ and $\theta_C\equiv -\theta_C$).
Thus all equivalence classes of UPBs are contained in the
region $\theta_A,\theta_B,\theta_C\in (0,\pi)$.
Although it is not necessary for the following discussion,
in Appendix we prove that all triples $(\theta_A,\theta_B,\theta_C)$
inside this region represent different equivalence classes.

By definition, an orthogonal complement
of a spanning space $\calH_S$ of any UPB $S$ 
does not contain product vectors. The set of product vectors
in the space $\calH_S$ itself however is not empty. It always contains
at least $n$ vectors --- the members of UPB.
Generally, the space $\calH_S$ contains more than $n$
product vectors
(some examples are given in Conclusion section).
Amazingly, in the system of three qubits the members of UPB
are the only product vectors in $\calH_S$, 
which significantly simplify analysis of convertibility issues.
 
\begin{lemma}\label{lemma:no-product-vectors}
Let $S$ be an arbitrary UPB for three qubits. The basis members
$|S_j\ra$, $j=1,2,3,4$ are the only product vectors in
the spanning space $\calH_S$. 
\end{lemma}
\begin{proof}
Let $S=\{|S_j\ra\}$ be a UPB with four members given by Eq.~(\ref{canon_UPB}).
We will prove more strong statement, namely, that
the basis members $|S_j\ra$, $j=1,2,3,4$ are the only 
product vectors in $\calH_S$ with respect to any {\it bipartite} cut of three
qubits. Consider for example a cut $ABC=A\cup BC$.
Suppose that for some $|\Psi_A\ra\in \CC^2$ and
some $|\Psi_{BC}\ra\in \CC^2\otimes\CC^2$ there exists
a decomposition
\be
\label{A+BC}
|\Psi_A\ra\otimes|\Psi_{BC}\ra=\sum_{j=1}^4 \alpha_j |S_j\ra,
\quad \sum_{j=1}^4 |\alpha_j|^2=1.
\ee
Tracing out the qubits B and C we get:
\be
\label{12+34}
|\Psi_A\ra\la\Psi_A| = \rho_{12} + \rho_{34},
\ee
where 
\[
\rho_{12}= \left( \ba{cc} 
|\alpha_1|^2 & \alpha_1\bar{\alpha}_2 \la B|0\ra \la C|0\ra \\
\bar{\alpha}_1\alpha_2 \la 0|B\ra \la 0|C\ra & |\alpha_2|^2\\
\ea\right),
\]
and
\[
\rho_{34}= \left( \ba{cc} 
|\alpha_3|^2 & \alpha_3\bar{\alpha}_4 \la B^\perp|1\ra \la 1|C^\perp\ra \\
\bar{\alpha}_3\alpha_4 \la 1|B^\perp\ra \la C^\perp|1\ra &
|\alpha_4|^2 \\ \ea \right).
\]
Here $\rho_{12}$ is represented by its matrix elements in $\{|0\ra,|1\ra\}$
basis, while $\rho_{34}$ 
is represented by its matrix elements in $\{|A\ra,|A^\perp\ra\}$
basis. Since $\rho_{12}$ and $\rho_{34}$ are non-negative
Hermitian operators, the equality in Eq.~(\ref{12+34}) is possible
only if $\rho_{12}$ and $\rho_{34}$ are both proportional to the
projector $|\Psi_A\ra\la\Psi_A|$. In particular it implies that
$\det{\rho_{12}}=0$ and $\det{\rho_{34}}=0$.
According to Lemma~\ref{lemma:UPBs} the bases $\{|B\ra,|B^\perp\ra\}$
and $\{|C\ra,|C^\perp\ra\}$
do not coincide with the basis $\{|0\ra,|1\ra\}$.
It means that the determinant
\[
\det \rho_{12} = |\alpha_1\alpha_2|^2 \left(
1 - |\la B|0\ra|^2 |\la C|0\ra|^2 \right)
\]
equals zero only if $\alpha_1\alpha_2=0$.
Analogously, $\det{\rho_{34}}=0$ only if $\alpha_3\alpha_4=0$.
Since the basis $\{|A\ra,|A^\perp\ra\}$ does not
coincide with $\{|0\ra,|1\ra\}$,
the equality in Eq.~(\ref{12+34})
is possible only if at most one of the coefficients
$\alpha_1$,  $\alpha_2$, $\alpha_3$, $\alpha_4$ is nonzero.
Thus the state $|\Psi_A\ra\otimes|\Psi_{BC}\ra$ coincides
with one of the basis members.
\end{proof}

\begin{cor}
UPBs $S$ and $T$ are equivalent iff $\rho_S$ and $\rho_T$
are related by local unitaries.
\end{cor}
Indeed, suppose that $\rho_T=U\rho_S U^\dag$, for some
unitary product operator $U=U_A\otimes U_B\otimes U_C$.
Then $U\calH_S=\calH_T$ and thus vectors $U|S_j\ra$,
$j=1,2,3,4$ are some product vectors in $\calH_T$.
Lemma~\ref{lemma:no-product-vectors} says that
these vectors are $|T_j\ra$, up to some permutation.
Thus $S\sim T$. The reverse statement is trivial. 

\begin{cor}
If UPBs $S$ and $T$ are not equivalent, they can not be
converted to each other by local filtering operations.
\end{cor}
To prove this corollary we will need two facts pointed out in 
Ref.~\cite{VDD01}, namely: 1)~An orbit of any
mixed state under LFO (or a closure of
this orbit) contains a {\it normal form}, i.e. a state
which has all one-particle marginals proportional to the identity.
2)~Inside each orbit there is a unique (up to local unitaries)
normal form. From the representation Eq.~(\ref{canon_UPB}) it is
clear that a state $\rho_S$ associated with an arbitrary UPB $S$
automatically comes in a normal form. Now the statement we
need follows from the previous corollary and 
the two facts mentioned above. To ivestigate convertibility
under general LOCC and approximate convertibility this results
however is not strong enough. We will need one more lemma.

\begin{lemma}\label{lemma:unitary}
Let $X=X_A\otimes X_B\otimes X_C$ be a non-degenerated 
(full rank) factorized
3-qubit operator such that 
\[
X\cdot \calH_S=\calH_T
\]
for some UPBs $S$ and $T$. Then $S\sim T$ and $X$ is proportional to
a unitary operator: $X=rU$, $r\in \RR$, $UU^\dag=I$.
\end{lemma}
\begin{proof}
Obviously, multiplying $X$ on factorized unitary operators on the left
and on the right is equivalent to a choice of representatives
in equivalence classes of $S$ and $T$. So we can assume that
\begin{itemize}
\item $X_A$, $X_B$, and $X_C$ are Hermitian operators
\item Members of the UPB $S$ are given by Eq.~(\ref{canon_UPB}).
\end{itemize}
Members of the UPB $T$ will be denoted as
\beea\label{T}
|T_1\ra&=&|A_1\ra\otimes|B_1\ra\otimes|C_1\ra,\nn \\
|T_2\ra&=&|A_2\ra\otimes|B_2\ra\otimes|C_2\ra,\nn \\
|T_3\ra&=&|A_3\ra\otimes|B_3\ra\otimes|C_3\ra,\nn \\
|T_4\ra&=&|A_4\ra\otimes|B_4\ra\otimes|C_4\ra. 
\eea
If $|\psi\ra\in \calH_S$ is a product vector, then 
its image $X|\psi\ra\in \calH_T$ is also a product vector.
Combining Lemma~\ref{lemma:no-product-vectors} and the fact that $X$
is non-degenerate operator, we conclude that 
\be
\label{S2T}
X|S_j\ra=x_j|T_{\sigma(j)}\ra, \quad j=1,2,3,4,
\ee
for some coefficients $x_j\ne 0$ and some permutation $\sigma\in S_4$.
Let us concentrate on the qubit A. From Eq.~(\ref{S2T}) we infer
\beea
\label{S2T'}
X_A|0\ra&=&a_1 |A_{\sigma(1)}\ra,\nn \\
X_A|1\ra&=&a_2 |A_{\sigma(2)}\ra,\nn \\
X_A|A\ra&=&a_3 |A_{\sigma(3)}\ra,\nn \\
X_A|A^\perp\ra&=&a_4 |A_{\sigma(4)}\ra,
\eea
where $a_j$ are some non-zero coefficients. 
For each $j=1,2,3,4$ choose some unital vector  $|A_j^\perp\ra$
orthogonal to $|A_j\ra$.
Taking into account the fact
that $X_A$ is Hermitian and non-degenerated, we can rewrite
Eq.~(\ref{S2T'}) as
\beea
\label{S2T''}
X_A|A_{\sigma(1)}^\perp\ra&=&b_1|1\ra,\nn \\
X_A|A_{\sigma(2)}^\perp\ra&=&b_2|0\ra,\nn \\
X_A|A_{\sigma(3)}^\perp\ra&=&b_3|A^\perp\ra,\nn \\
X_A|A_{\sigma(4)}^\perp\ra&=&b_4|A\ra,
\eea
for some non-zero coefficients $b_j$. 
Applying the classificaltion Lemma~\ref{lemma:UPBs} to the UPB $T$
we conclude that the
family $\{|A_j^\perp\ra\}$ and the family $\{|A_j\ra\}$
coincide up to permutation of the elements and
some phase adjustment for each element. 
Moreover, from this Lemma we can learn that an
orthogonality graph for the family
$\{|A_j\ra\}$ must be one of three graphs shown on FIG.~2.\par
\smallskip

\begin{figure}[ht]
\includegraphics[scale=0.5]{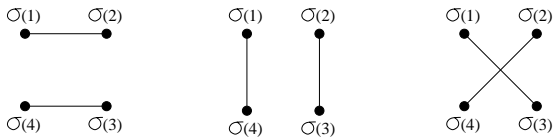}
\caption{Possible orthogonality graphs for
the family $\{|A_j\ra\}$.}
\end{figure}\noindent
(We choose an ordering of vertices which depends upon $\sigma$,
since this ordering appears in Eq.~(\ref{S2T'},\ref{S2T''}).)
Whatever graph is chosen, the equations Eq.~(\ref{S2T'},\ref{S2T''})
guarantee that an operator $(X_A)^4$ is diagonal both in the
basis $\{|0\ra,|1\ra\}$ and the basis $\{|A\ra,|A^\perp\ra\}$.
By definition, these bases do not coincide. Thus $(X_A)^4$
is proportional to the identity, $(X_A)^4=x_A I$, for some
coefficient $x_A$. Since $X_A$ is Hermitian,
this is possible only if  $X_A$ is proportional
to a unitary operator. Applying the same arguments to
the qubits B and C, we conclude that $X$ is proportional to
a unitary operator. Finally, Lemma~\ref{lemma:no-product-vectors}
implies that $S\sim T$. 
\end{proof}

\section{\label{sec:main} Exact convertibility of bound entangled states}
Consider a pair of 3-qubit mixed states $\rho_S$ and $\rho_T$
associated with a pair of {\it non-equivalent} UPBs $S$ and $T$
respectively. Non-equivalence assumption implies that we can not
convert $\rho_S$ into $\rho_T$ {\it exactly} by LFO,
see the corollary to Lemma~\ref{lemma:no-product-vectors}.
Now we are in position to prove more strong statement, namely
that $\rho_S$ can not be converted into $\rho_T$ exactly by a
separable superoperators.
To simplify notations, let us denote
$\Omega_f$ a set of 3-qubit product operators, i.e.
\[
\Omega_f=\{ X=X_A\otimes X_B\otimes X_C, \quad X_A, X_B, X_C\in \LL(\CC^2)\}.
\]
\begin{lemma}\label{lemma:exact}
Let $\{X_l\in \Omega_f\}_{l=1,\ldots,D}$ be a family of product operators
such that 
$\sum_{l=1}^D X_l^\dag X_l \le I$.
Suppose that for some UPBs $S$, $T$ and real number $p>0$
one has the equality
\be\label{exact}
\sum_{l=1}^D X_l \, \rho_S \, X_l^\dag = p\rho_T.
\ee
Then the UPBs $S$ and $T$ are equivalent. 
\end{lemma}\noindent
{\it Remarks:} 1) The factor $p$ can be regarded as a
probability for conversion to succeed, since
\[
p=\tr(\sum_{l=1}^D X_l^\dag X_l \rho_S)\in [0,1].
\]
2) Without loss of generality we can choose $D$ as
a real dimension of a space of all superoperators on three qubits,
which is $D=2^{13}$. Indeed, superoperators $E(\rho)=\sum_l X_l\rho X_l^\dag$
discussed in the lemma constitute a compact convex subset in the
linear space of all superoperators on three qubits.

\begin{proof}
Denote 
\[
\rho_l=X_l\, \rho_S \, X_l^\dag.
\]
Since $\rho_l$ is a positive semidefinite
operator, the equality Eq.~(\ref{exact}) tells us that
\be\label{inclusion}
\range{\rho_l}\subseteq \range{\rho_T},\quad
l=1,\ldots, D,
\ee
or, equivalently,
\be\label{inclusion1}
X_l \cdot \range{\rho_S} \subseteq \range{\rho_T},
\ee
where $\range{\rho}$ is a range of the operator $\rho$.
Suppose that for some $l$ 
we have $\rho_l\ne 0$ and the operator $X_l$
is degenerated.  Let us show that this assumption leads
to a contradiction.
By definition, $X_l$ has a product
form, $X_l=X_A\otimes X_B\otimes X_C$, so  at least one 
of the factors $X_A$, $X_B$, $X_C$ has a rank one. 
Consider the case $\rk{X_A}=1$, i.e. $X_A=|\psi\ra\la\phi|$
for some vectors $|\psi\ra,|\phi\ra\in \CC^2$.
Then
\beea\label{PPT}
\rho_l &=& |\psi\ra\la \psi|\otimes \rho_{BC},\nn \\
\rho_{BC}&=&X_B\otimes X_C \, \la \phi|\rho_S|\phi\ra\,
X_B^\dag\otimes X_C^\dag.
\eea
Since $\rho_S$ is a positive partial traspose (PPT) state,
the same does $\rho_l$. Then from the equality Eq.~(\ref{PPT})
we infer that $\rho_{BC}$ is a two-qubit PPT state.
But according to Peres criteria~\cite{Peres96,HHH96}
it implies that $\rho_{BC}$ is
separable.  Let us choose some product vector
$|\psi_B\ra\otimes|\psi_C\ra$ from its range.
Then a product vector $|\psi\ra\otimes|\psi_B\ra\otimes|\psi_C\ra$
belongs to the range of $\rho_l$
and thus to the range of $\rho_T$, see Eq.~(\ref{inclusion}).
Since $\rho_T$ has no product vectors
in its range, we conclude that for each $l$
either $X_l$ is non-degenerated or $\rho_l=0$.

Let us focus on some $l$ with $\rho_l\ne 0$. 
According to classification Lemma~\ref{lemma:UPBs},
the spaces $\range{\rho_S}$ and $\range{\rho_T}$
have the same dimension.
Since $X_l$ is non-degenerated,  Eq.~(\ref{inclusion1})
actually means that $X_l\cdot\range{\rho_S}=\range{\rho_T}$,
or equivalently, that
\be\label{impossible}
X_l^\dag \cdot\calH_T = \calH_S.
\ee
According to Lemma~\ref{lemma:unitary} this is possible
only if $S\sim T$.
\end{proof}
Since stochastic LOCC transformations are described by
separable superoperators,
we have proved the following theorem:
\begin{theorem}\label{theorem:exact}
Let $S$ and $T$ be unextendible product bases 
for three qubits. Consider bound entangled states
$\rho_S$ and $\rho_T$ associated with $S$ and $T$.
If some LOCC transformation maps $\rho_S$
into $\rho_T$ with a non-zero probability, then
the states $\rho_S$ and $\rho_T$ coincide up to local
unitary transformation.
\end{theorem}

\section{\label{sec:approx1} Approximate convertibility: simplified
scenario}

Consider a three-qubit UPB $S=\{|S_j\ra\}$ with the members given by
Eq.~(\ref{intro1}). Let $T$ be another UPB which is not
equivalent to $S$. In this section we will study a simplified
version of the stochastic approximate 
convertibility problem. Namely, we will consider
approximate convertibility by local filtering operations, 
i.e. transformations like
\be\label{rho}
\rho_S\to \rho=\frac{X \rho_S X^\dag}{p_S[X]}, \quad
p_S[X]\equiv\tr(X \rho_S X^\dag).
\ee
where  $X\in \Omega_f$ is a product operator such that
$p_S[X]>0$. Our goal is to prove that there exists a
finite precision $\ep>0$ such that all states $\rho$ which
can appear in Eq.~(\ref{rho}) 
lie outside $\ep$-neighborhood
of $\rho_T$ (we use fidelity to quantify the distance).
Denote a set of achievable states $\rho$ as $M_S$:
\be\label{MS}
M_S= \left\{ \rho= \frac{X \rho_S X^\dag}{p_S[X]} \ : \, 
X\in \Omega_f,\quad  p_S[X]>0  \right\}.
\ee
The results of the previous section imply that $\rho_T\notin M_S$.
Note however that $M_S$ is not a compact set since $p_S[X]=0$ for
some product operators $X$ (for instance, $p_S[|S_j\ra\la S_j|]=0$
for any member of the UPB). Thus there might exist a
sequence of product operators $\{X_n\}_{n\ge 0}$,
$p_S[X_n]>0$,
such that the corresponding sequence  $\rho(X_n)$  converges to some operator
$\rho$ which does not belong to $M_S$. If $\rho=\rho_T$,
it would imply that stochastic convertibility with an arbitrary small
precision $\ep$ is possible (although the success probability may turn to
zero as $\ep \to 0$). Amazingly, this is not the case.
\begin{lemma}\label{lemma:simplified}
Let $S$ be an arbitrary UPB. 
Consider a set of achievable states $M_S$ as in Eq.~(\ref{MS}) and
an arbitrary convergent operator sequence 
$\{Y_n \in M_S\}_{n\ge 0}$. Denote
\[
\hat{Y} = \lim_{n\to \infty} Y_n.
\]
Then either $\hat{Y}\in M_S$ or $\hat{Y}$ is separable. 
\end{lemma}
\begin{proof}
Suppose that $\hat{Y}\notin M_S$.
Obviously, in the definition of $M_S$ one suffices to consider 
normalized operators $X$. Let us agree that 
\[
X=X_A\otimes X_B\otimes X_C, \quad ||X_A||=||X_B||=||X_C||=1.
\]
Let us choose an arbitrary sequence of normalized operators
$\{X_n \in \Omega_f\}_{n\ge 0}$, $p_S[X_n]>0$, such that
$Y_n=X_n\rho_S X_n^\dag/p_S[X_n]$.
Without lose of generality we can assume that the sequence
$\{X_n\}$ is also convergent (since $X_n$ are taken from
a bounded manifold, we can always extract a convergent
subsequence). Denote
\be\label{Xlimit}
\hat{X}=\lim_{n\to \infty} X_n=\hat{X}_A\otimes\hat{X}_B\otimes \hat{X}_C.
\ee
Note that  $p_S[\hat{X}]=0$, since otherwise $\hat{Y}\in M_S$.
It means that $\hat{X}\rho_S\hat{X}^\dag=0$,
or, equivalently, $\rho_S \hat{X}^\dag=0$ which implies 
$\image{\hat{X}^\dag}\subseteq \kernel{\rho_S}$. Therefore
\[
\image{\hat{X}_A^\dag}\otimes
\image{\hat{X}_B^\dag}\otimes \image{\hat{X}_C^\dag}
\subseteq \calH_S.
\]
But according to Lemma~\ref{lemma:no-product-vectors}
the only product vectors in $\calH_S$ are the members of $S$.
Thus $\hat{X}=|a,b,c\ra\la S_j|$ for some $j\in [1,4]$
and some one-qubit normalized states $|a\ra$, $|b\ra$, $|c\ra$
Without loss of generality, we can assume that $j=1$, i.e. 
\be\label{Xlimit'}
\hat{X}_A=|a\ra\la 0|, \quad 
\hat{X}_B=|b\ra\la 0|, \quad 
\hat{X}_C=|c\ra\la 0|,
\ee
(see Eq.~(\ref{canon_UPB})).
The elements of the sequence $\{X_n\}_{n\ge 0}$ can always be
written as
\beea
X_n&=&X_{A,n}\otimes X_{B,n}\otimes X_{C,n},\nn \\
X_{A,n}&=&|a_n\ra\la 0| + |\alpha_n\ra\la 1|, \nn \\
X_{B,n}&=&|b_n\ra\la 0| + |\beta_n\ra\la 1|, \nn \\
X_{C,n}&=&|c_n\ra\la 0| + |\gamma_n\ra\la 1|, 
\eea
where $|a_n\ra$, $|b_n\ra$, $|c_n\ra$ and
$|\alpha_n\ra$, $|\beta_n\ra$, $|\gamma_n\ra$ are some
one-qubit states. The requirements Eq.~(\ref{Xlimit},\ref{Xlimit'})
translate into
\[
\lim_{n\to \infty} |a_n\ra = |a\ra,
\quad
\lim_{n\to \infty} |b_n\ra = |b\ra,
\quad
\lim_{n\to \infty} |c_n\ra = |c\ra,
\]
and
\[
\lim_{n\to \infty} \la\alpha_n|\alpha_n\ra =
\lim_{n\to \infty} \la\beta_n|\beta_n\ra =
\lim_{n\to \infty} \la\gamma_n|\gamma_n\ra = 0.
\]
Denote
\be
Y_n = \frac{X_n \rho_S X_n^\dag}{p_S[X_n]}.
\ee
To compute a limit of $Y_n$ for $n\to \infty$
we can keep only the leading terms in the
expression for $X_n$ (see a comment below), namely 
\beea
X_n &\approx& X_n^{(0)} + X_n^{(1)},\nn \\
X_n^{(0)}&=&|a_n, b_n, c_n\ra\la 0,0,0|, \nn \\
X_n^{(1)}&=&|\alpha_n,b_n,c_n\ra\la 1,0,0| +
|a_n,\beta_n,c_n\ra\la 0,1,0|
\nn \\
&& {} +
|a_n,b_n,\gamma_n\ra\la 0,0,1|.
\eea
Obviously, $X_n^{0}$ does not yield any contribution to $Y_n$,
so that 
\[
X_n\rho_S X_n^\dag \approx X_n^{(1)} \rho_S {X_n^{(1)}}^\dag.
\]
Now we can substitute $a_n, b_n, c_n$ in the expression for
$X_n^{(1)}$ by $a,b,c$ respectively, since
we would like to keep only leading order terms.
Using the list of the UPB's members Eq.~(\ref{canon_UPB}) one can
easily check that  matrix elements 
$\la 1,0,0|\rho_S|0,1,0\ra$, 
$\la 1,0,0|\rho_S|0,0,1\ra$  (and their cyclic permutations)
vanish. Therefore we arrive to 
\beea\label{approx1}
X_n^{(1)}\rho_S {X_n^{(1)}}^\dag \approx && \nn \\
\approx
\la 1,0,0|\rho_S|1,0,0\ra \cdot |\alpha_n,b,c\ra\la \alpha_n,b,c|  \nn \\
+ \la 0,1,0|\rho_S|0,1,0\ra \cdot |a,\beta_n,c\ra\la a,\beta_n,c|  \nn \\
+ \la 0,0,1|\rho_S|0,0,1\ra \cdot |a,b,\gamma_n\ra\la a,b,\gamma_n|. 
\eea
Note that this is a separable state and that
the matrix elements 
\[
\la 1,0,0|\rho_S|1,0,0\ra, \,
\la 0,1,0|\rho_S|0,1,0\ra, \,
\la 0,0,1|\rho_S|0,0,1\ra,
\]
are strictly positive, since 
Lemma~\ref{lemma:no-product-vectors} tells us that
the states $|1,0,0\ra$,
$|0,1,0\ra$, and $|0,0,1\ra$ do not belong to $\calH_S$.
It justifies that the leading terms in the expansion for
$X_n\rho_S X_n^\dag$ do not vanish and all terms we have
disregarded are indeed small compared with the terms we keep.
In particular,
\beea\label{approx2}
\tr(X_n^{(1)} \rho_S {X_n^{(1)}}^\dag) \approx && \nn \\
\la 1,0,0|\rho_S|1,0,0\ra \la\alpha_n|\alpha_n\ra \nn \\
+
\la 0,1,0|\rho_S|0,1,0\ra \la\beta_n|\beta_n\ra \nn \\
+
\la 0,0,1|\rho_S|0,0,1\ra \la \gamma_n|\gamma_n\ra.
\eea
We conclude that
\be
\hat{Y}\equiv \lim_{n\to\infty} Y_n = \lim_{n\to \infty}
\frac{X_n^{(1)}\rho_S {X_n^{(1)}}^\dag }
{\tr\left(X_n^{(1)} \rho_S {X_n^{(1)}}^\dag\right)},
\ee
where approximations Eq.~(\ref{approx1},\ref{approx2}) should be
substituted into nominator and denominator.
But from Eq.~(\ref{approx1}) we infer that $\hat{Y}$ is separable. 
\end{proof}
Denote $\overline{M}$ a closure of the set $M$. It is a compact set.
A fidelity $F(\rho,\rho_T)$ is a continuous function of
a state $\rho$. According to  Lemma~\ref{lemma:exact},
$F(\rho,\rho_T)<1$ for all $\rho\in M_S$. Lemma~\ref{lemma:simplified}
implies that $F(\rho,\rho_T)<1$ even for all $\rho\in \overline{M_S}$.
Thus there exists a finite precision $\ep>0$ such that
\[
F(\rho,\rho_T)\le 1-\ep, \quad \mbox{for any} \quad
\rho\in M_S.
\]
Of course the precision $\ep$ may depend upon $S$ and $T$,
in particular $\ep\to 0$ as $S$ turns to $T$.

\section{\label{sec:approx2} Approximate convertibility in a general case}
We now are prepared to analyse approximate
convertibility by arbitrary completely positive separable superoperators. 
Such operators correspond to probabilistic mixtures of LFO,
so we can exploit the results of section~\ref{sec:approx1}.
A proof of Theorem~\ref{theorem:main} is contained in the following lemma.

\begin{lemma}
Let $S$ and $T$ be non-equivalent UPBs.
Suppose $\{X_l\in \Omega_f\}_{l=1,\ldots,D}$ is a family of product
operators such that 
$p_S[X_l]\equiv \tr(X_l\rho_S X_l^\dag)>0$ for all $l$.
Denote $p_S=\sum_{l=1}^D p_S[X_l]$. There exist a finite
precision $\ep=\ep(S,T)>0$ depending upon $S$ and $T$ only,
such that
\be
F\left(\rho_T,\frac1{p_S} \sum_{l=1}^D X_l \rho_S X_l^\dag\right)\le 
1-\ep(S,T).
\ee
\end{lemma}
\begin{proof}
We start from introducing an auxiliary proximity measure
between $\rho_T$ and the final state. Let $|T_j\ra$,
$j=1,2,3,4$ be the members of the UPB $T$. Consider a functional
\be\label{f_T}
f_T(\rho)= \sum_{j=1}^4 \la T_j|\rho| T_j\ra.
\ee
It is clear that 
\be\label{f=0}
f_T(\rho)=0 \quad  \mbox{iff} \quad \range{\rho}\subseteq \range{\rho_T}.
\ee
Thus the equality $f_T(\rho)=0$ is necessary for $\rho=\rho_T$.
Define $\rho$ as
\[
\rho=\frac1{p_S} \sum_{l=1}^D X_l \rho_S X_l^\dag.
\]
Let us first prove that there exist a finite precision $\delta>0$
depending upon $S$ and $T$ only, such that
\be\label{important-estimate}
f_T(\rho)\ge \delta.
\ee
Using linearity of $f_T$ we can write
\beea\label{f_T1}
f_T(\rho)&=&\sum_{l=1}^D \frac{p_S[X_l]}{p_S} f_T(Y_l),\nn \\
Y_l&\equiv& \frac{X_l \rho_S X_l^\dag}{p_S[X_l]} \in M_S,
\eea
see Eq.~(\ref{MS}).
Using the same arguments as in the proof of Lemma~\ref{lemma:exact}
one can easily show that an inclusion $\range{Y_l}\subseteq \range{\rho_T}$
is impossible
for any $Y_l\in M_S$. From Eq.~(\ref{f=0}) we infer that
$f_T(Y_l)>0$ for any $Y_l\in M_S$. Moreover, $f_T(Y)>0$ for any
$Y$ belonging to a closure of the set $M_S$. Indeed, Lemma~\ref{lemma:simplified}
tells us that such $Y$ either belongs to $M_S$
or is separable. In the latter case the inclusion
$\range{Y}\subseteq \range{\rho_T}$ is impossible, since 
$\range{\rho_T}$ is free of product vectors. Summarizing, there 
exists a finite precision $\delta$, such that
\be
f_T(Y)\ge \delta, \quad \mbox{for any} \quad Y\in M_S.
\ee
Returning to Eq.~(\ref{f_T1}) we immediately get
the estimate Eq.~(\ref{important-estimate}).

To conclude the proof we need to turn inequality Eq.~(\ref{important-estimate})
into an upper bound on fidelity $F(\rho_T,\rho)$. It can be done
as follows. We start from the standard definition of fidelity:
\be\label{fidelity}
F(\rho_T,\rho)=\tr\left[
\left(\sqrt{\rho_T} \, \rho\, \sqrt{\rho_T}\right)^\frac12\right].
\ee
Introduce orthogonal projectors $P_T$ and $P_T^\perp$ onto
the spanning space $\calH_T$ and its orthogonal complement
respectively. By definition, $P_T+ P_T^\perp= I$.
 Then  $\rho_T=(1/4) P_T^\perp$ and thus
\be\label{fidelity1}
F(\rho_T,\rho)=\frac12 \tr \sqrt{P_T^\perp \rho P_T^\perp}.
\ee
Note that $f_T(\rho)=\tr (\rho P_T)$.
From an identity
\[
\tr \left( P_T^\perp \rho P_T^\perp \right) +
\tr \left( P_T \rho P_T \right) =1
\]
and from the estimate Eq.~(\ref{important-estimate}) we get
\be
\tr \left( P_T^\perp \rho P_T^\perp\right) \le 1-\delta .
\ee
Taking into account that the operator $P_T^\perp \rho P_T^\perp$
has a rank at most four and
applying Cauchy-Schwarz inequality we arrive to
\be
\tr \sqrt{P_T^\perp \rho P_T^\perp} \le
\sqrt{ 4 \tr \left( P_T^\perp \rho P_T^\perp\right) }
\le 2 \sqrt{1-\delta}.
\ee
Substituting it to Eq.~(\ref{fidelity1}) gives us
\be
F(\rho_T,\rho) \le \sqrt{1-\delta}\le 1 - \frac{\delta}2.
\ee
We have proved the lemma.
\end{proof}

\section{\label{sec:conclusion} Conclusion}
We have studied local convertibility of three-qubit mixed states
associated with unextendible product bases.
A complete classification of three-qubit UPBs is suggested.
This family of UPBs is shown to have some nice mathematical
properties which allow to investigate convertibility question
completely. We proved that for any non-equivalent UPBs $S$ 
and $T$ the stochastic approximate conversion of the 
associated states $\rho_S$ and $\rho_T$ is impossible.
 
It would be interesting to apply our method to UPBs in some
other low-dimension systems. The system of two qutrits 
$\CC^3\otimes \CC^3$ is of particular interest, since
for this system a complete UPBs classification has been already 
found in Ref.~\cite{DMSST00}.
 It is known that all two-qutrit UPBs are characterized
by the same orthogonality graphs and consist of five members.
Unfortunately a set of product vectors in a spaning space
of a UPB is generally larger than a set of the UPB's members.
We have checked it for two particular two-qutrit UPBs, called
in Ref.~\cite{BDMSST99} as "Tiles" and "Pyramid". Using the designations of
this reference, the extra product vectors in the spaning spaces
of "Pyramid" and "Tiles" respectively can be written as 
\[
|0\ra\otimes |0\ra \quad \mbox{and} \quad
\frac19 (2|0\ra-|1\ra+2|2\ra)\otimes (2|0\ra-|1\ra+2|2\ra).
\]
Exploiting the symmetry of these particular UPBs one can
show that there are exactly six product vectors in their
spaning spaces. However
the presence of extra product vectors as well as three-dimensional
geometry makes the convertibility analysis very complicated.

\begin{acknowledgments}
We would like to thank Guifre Vidal for supporting ideas
which significantly simplify the proof in Section~\ref{sec:main}.
We also acknowledge useful conversations with Patrick Hayden,
Alexei Kitaev, and Federico Spedalieri.
\end{acknowledgments}

\appendix*
\section{}
In this section we prove that all UPBs given by Eq.~(\ref{canon_UPB},\ref{ABC})
with parameters $\theta_A,\theta_B,\theta_C\in (0,\pi)$ are not equivalent
to each other in the sense of Definition~\ref{def:equiv}.
Consider  UPBs $S=\{|S_j\ra\}$ and $S'=\{|S_j'\ra\}$ with
the members
\beea
|S_1\ra&=&|0\ra\otimes|0\ra\otimes|0\ra, \nn \\
|S_2\ra&=&|1\ra\otimes|B\ra\otimes|C\ra, \nn \\
|S_3\ra&=&|A\ra\otimes|1\ra\otimes|C^\perp\ra,\nn  \\
|S_4\ra&=&|A^\perp\ra\otimes|B^\perp\ra\otimes|1\ra, 
\eea
and 
\beea
|S_1'\ra&=&|0\ra\otimes|0\ra\otimes|0\ra, \nn \\
|S_2'\ra&=&|1\ra\otimes|B'\ra\otimes|C'\ra, \nn \\
|S_3'\ra&=&|A'\ra\otimes|1\ra\otimes|{C'}^\perp\ra,\nn \\
|S_4'\ra&=&|{A'}^\perp\ra\otimes|{B'}^\perp\ra\otimes|1\ra.
\eea
Since the overall phase of the vectors is not important,
the parameterization Eq.~(\ref{intro2}) is equivalent to
\beea\label{ABC1}
|A\ra&=&\cos{(\theta_A/2)}|0\ra + \sin{(\theta_A/2)} |1\ra, \nn \\
|B\ra&=&\cos{(\theta_B/2)}|0\ra + \sin{(\theta_B/2)} |1\ra,\nn \\
|C\ra&=&\cos{(\theta_C/2)}|0\ra + \sin{(\theta_C/2)} |1\ra,
\eea
and
\beea\label{ABC2}
|A'\ra&=&\cos{(\theta_A'/2)}|0\ra + \sin{(\theta_A'/2)} |1\ra,\nn \\
|B'\ra&=&\cos{(\theta_B'/2)}|0\ra + \sin{(\theta_B'/2)} |1\ra,\nn \\
|C'\ra&=&\cos{(\theta_C'/2)}|0\ra + \sin{(\theta_C'/2)} |1\ra.
\eea
Suppose that 
\[
U_A\otimes U_B\otimes U_C |S_j\ra = |S_{\sigma(j)'}\ra
\]
for some one-qubit unitary operators $U_A$, $U_B$, $U_C$
and permutation $\sigma\in S_4$.
If $\sigma(1)=1$ then orthogonality implies  $\sigma(j)=j$
for all $j$. It means that $|\la 0|A\ra|=|\la 0 |A'\ra|$,
$|\la 0|B\ra|=|\la 0 |B'\ra|$, and $|\la 0|C\ra|=|\la 0 |C'\ra|$.
From Eq.~(\ref{ABC1}) and Eq.~(\ref{ABC2})
we have $|\cos{(\theta_A/2)}|=|\cos{(\theta_A'/2)}|$.
Since $\theta_A,\theta_A'\in (0,\pi)$ we conclude that
$\theta_A'=\theta_A$.  Analogously one gets
$\theta_B'=\theta_B$ and $\theta_C'=\theta_C$.

If $\sigma(2)=1$ then orthogonality implies
$\sigma(2)=1$, $\sigma(3)=4$, and $\sigma(4)=3$.
It means that $|\la 0|A\ra|=|\la 1 |{A'}^\perp\ra|$,
$|\la 0|B\ra|=|\la 0 |B'\ra|$, and $|\la 0|C\ra|=|\la 0 |C'\ra|$.
Thus $(\theta_A,\theta_B,\theta_C)$ and $(\theta_A',\theta_B',\theta_C')$
coincide.

If $\sigma(3)=1$ then orthogonality implies
$\sigma(1)=3$, $\sigma(2)=4$, and $\sigma(4)=2$.
It means that $|\la 0|A\ra|=|\la 0 |A'\ra|$,
$|\la 0|B\ra|=|\la 1 |{B'}^\perp\ra|$,
and $|\la 0|C^\perp\ra|=|\la 0 |{C'}^\perp\ra|$.
Thus $(\theta_A,\theta_B,\theta_C)$ and $(\theta_A',\theta_B',\theta_C')$
coincide.

Finally, if $\sigma(4)=1$ then orthogonality implies
$\sigma(1)=4$, $\sigma(2)=3$, and $\sigma(3)=2$.
It means that $|\la 0|A^\perp\ra|=|\la 0 |{A'}^\perp\ra|$,
$|\la 0|B^\perp\ra|=|\la 0 |{B'}^\perp\ra|$,
and $|\la 0|C\ra|=|\la 1 |{C'}^\perp\ra|$.
Thus $(\theta_A,\theta_B,\theta_C)$ and $(\theta_A',\theta_B',\theta_C')$
coincide.


\end{document}